\documentstyle[aps,prb,twocolumn,floats,epsf]{revtex}

\input{psfig}
\begin{document}
\title{The Role of Spin-Dependent Interface Scattering in Generating
Current-Induced Torques in Magnetic Multilayers}

\author{Xavier Waintal, Edward B. Myers, Piet W. Brouwer, and D.
C. Ralph}
\address
{Laboratory of Atomic and Solid State Physics, Cornell University,
Ithaca NY 14853, USA
\\ {\rm (\today)}
\medskip ~~\\ \parbox{14cm}{\rm	We present a calculation of 
current-induced torques in metallic magnetic
multilayers derived from the spin-dependent transmission and reflection
properties of the magnetic layers. A scattering formalism
is employed to calculate the torques in a magnetic-nonmagnetic-magnetic
trilayer, for currents perpendicular to the layers, in both the ballistic
and diffusive regimes.
}}

\maketitle
   
\section{Introduction}
Stacks of alternating ferromagnetic and nonmagnetic metal layers
exhibit giant magnetoresistance (GMR), because their electrical resistance
depends strongly on whether the moments of adjacent magnetic layers are
parallel or antiparallel.  This effect has allowed the development of new
kinds of field-sensing and magnetic memory devices.\cite{ibm}
The cause of the GMR effect is that conduction electrons are 
scattered more strongly by
a magnetic layer when their spins lie antiparallel to the layer's magnetic
moment than when their spins are parallel to the moment. Devices with
moments in adjacent magnetic layers aligned antiparallel thus have a larger
overall
resistance than when the moments are aligned parallel, giving rise to GMR.
This paper discusses the converse effect: just as the orientations of
magnetic moments can affect the flow of electrons, then by Newton's third
law, a polarized electron current  scattering from a magnetic layer can
have a reciprocal effect on the moment of the layer.
As proposed by Berger\cite{berger} and
Slonczewski,\cite{slon1} an electric current passing
perpendicularly through a magnetic multilayer may exert a torque
on the moments of the magnetic layers. This effect which is
known as ``spin transfer'',\cite{deceptively} may, at sufficiently high
current densities, alter the magnetization state. It is a separate mechanism
from the effects of current induced magnetic fields. 
Experimentally, spin-current-induced
magnetic excitations such as spin-waves,\cite{tsoi,louie,myers,katine}
and stable magnetic reversal,\cite{myers,katine} have been observed
in multilayers, for current densities greater than $10^7 A/cm^2$.

	The spin-transfer effect offers the promise of new kinds of magnetic
devices,\cite{slonrev} and serves as a new means to excite and to probe
the dynamics of magnetic moments
at the nanometer scale.\cite{tsoi2} In order to controllably utilize
these effects, however, it is necessary to achieve a better quantitative
understanding of these current-induced torques. Slonczewski has
presented a derivation of spin-transfer torques using a 1-D WKB approximation
with spin-dependent potentials,\cite{slon1} but his calculations only
take into account
electrons which are either completely transmitted or completely reflected by
the magnetic layers. For real materials the degree to which an electron
is transmitted through a magnetic/nonmagnetic interface depends sensitively on
the matching of the band structures across the interface.\cite{stiles,schep}
It is the goal of this paper to incorporate such band structure
information
together with the effect of multiple reflections between the
ferromagnetic layers,
into a more quantitative theory of the torques generated by
spin-transfer.
This could be done using the formalism of Brataas {\it et al},\cite{brataas}
which is based on kinetic equations for spin currents. Instead we
choose to employ 
a modified Landauer-B\"uttiker formalism, in which we model
the ferromagnetic layers as generalized spin-dependent scatterers. The
calculations
are carried out for a quasi-one dimensional geometry,
for which we derive formulas
for the torque generated on the magnetic layers when a current is
applied to the system, for either ballistic or diffusive non-magnetic
layers.
The main difference between our approach and Ref.~\onlinecite{brataas}
is that in our case, scattering in the normal layer is phase coherent,
whereas Ref.~\onlinecite{brataas} assumes phase relaxation.
However, in the case of a diffusive normal metal layer and for a large
number of transverse modes, the two approaches would give the same answer.

	The paper is organized as follows: in Section~\ref{1}, we
	present an intuitive picture (adapted from\cite{berger,slon1}) of how
spin-dependent scattering of a spin-polarized current produces a torque on
a magnetic element. Section~\ref{2} is devoted to the introduction
	of the scattering matrix formalism for the spin-flux. This
	formalism is then used in Section~\ref{3}
 	to calculate the torque in a Ferromagnet-Normal-Ferromagnet (FNF)
	system where the normal part is disordered (diffusive).
	Section~\ref{5} contain a discussion of the
	results.  Details of our calculation are presented in Appendix
	A. In  appendix B, we derive the torque for an FNF system where
transport in the normal layer is ballistic, rather than diffusive.

\section{Physical Idea}
\label{1}

In this section we will present a simple intuitive picture of the
physics behind the spin-transfer effect.  The connection between
current-induced spin-transfer torques and the spin-dependent
scattering that occurs when electrons pass through a
magnetic/nonmagnetic interface can be illustrated most simply by
considering the case of a spin-polarized current incident
perpendicularly on a single thin ferromagnetic layer F, as shown in
Figure 1.  The layer lies in the $y-z$ plane, with its magnetic moment
uniformly pointed in the $+z$ direction, and we assume that the
current is spin-polarized in the $z-x$ plane at an angle $\theta$ to
the layer moments.  The incoming electrons can therefore be considered
as a coherent linear superposition of basis states with spin in the +z
direction (amplitude $\cos(\theta/2)$) and -z direction (amplitude
$\sin(\theta/2)$).  For this initial discussion we will assume that
the layer is a perfect spin-filter, so that spins aligned with the
layer moments are completely transmitted through the layer, while
spins aligned antiparallel to the layer moment are completely
reflected.  For incident spins
polarized at an angle $\theta$, the average outgoing current will have
the relative weights $\cos^2(\theta/2)$ polarized in the $+z$ direction
and transmitted to the right and $\sin^2(\theta/2)$ polarized in the $-z$
direction and reflected to the left.  Consequently, both of the
outgoing electron spin fluxes (transmitted and reflected) lie along
the $z$ axis, while the incoming (incident) electron flux has a
component perpendicular to the magnetization, along the $x$ axis, with
magnitude proportional to $\sin\theta$. This $x$-component of angular
momentum must be absorbed by the layer in the process of
filtering the spins. Because the spin-filtering is ultimately governed
by the $s-d$ exchange interaction between the conduction electrons and
the magnetic moments of the layer, the angular momentum is imparted to
the layer moments and produces a torque on them.  This exchange
torque,\cite{bergerold} which is proportional to the electron current
through the layer and to $\sin\theta$, is in the direction to align
the moments with the polarization of the incident spin current.

\begin{figure}
\centerline{\psfig{figure=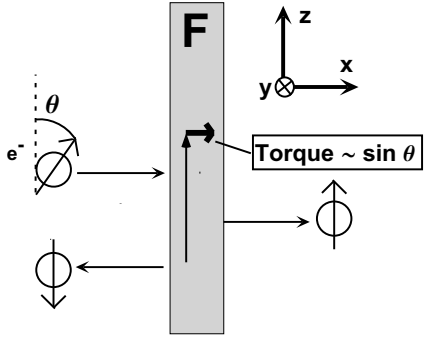,width=7.0cm}}
\vspace{3mm} \narrowtext
\caption{
\label{figure1}
Schematic of exchange torque generated by spin-filtering. Spin-polarized
electrons are incident perpendicularly on a thin ideal ferromagnetic layer.
Spin-filtering removes the component of spin angular momentum perpendicular
to the layer moments from the current; this is absorbed by the moments
themselves, generating an effective torque on the layer moments.}
\end{figure}

\begin{figure}[tbh]
\centerline{\psfig{figure=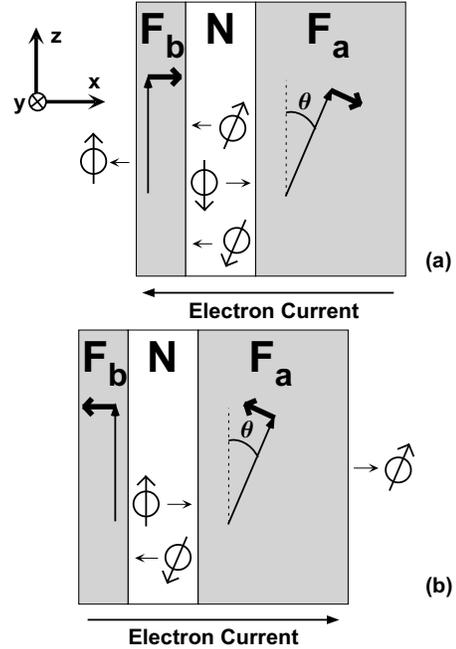,width=7.0cm}}
\vspace{3mm} \narrowtext
\caption[fig2]{
\label{figure2}
Qualitative picture of asymmetry of spin-transfer torque with respect
to current bias in a FNF junction. For left-going electrons (2a),
initially polarized by a magnetic layer F$_a$, the moments of layer F$_b$
experience a torque so as to align them with layer F$_a$. The electron
current reflected from layer F$_b$, in turn, exerts a torque on layer
F$_a$ so as to antialign it with the moment of layer F$_b$. Subsequent
reflections between the layers reduce but do not eliminate this
torque. If the current is reversed $(2b)$ the overall sign of the
torque is reversed, encouraging the moment of layer F$_b$ to align
antiparallel with layer F$_a$.}
\end{figure}

The symmetry of this model precludes any generation of torque
from the spin-filtering of a current of unpolarized electrons. To generate
the effect, then, a second ferromagnetic layer is needed to first spin-polarize
the current, see Fig.\ \ref{figure2}.  In that case,  spin angular momentum
is transferred from one layer to the flowing electrons and then from the
electrons to the second layer.  However, the  torques on the two layers
are not equal and opposite, as spin angular momentum carried by the
electrons can also flow
away from the layers to infinity, see Fig.\ \ref{figure2}.

The presence of this second layer has the additional effect of
allowing for multiple scattering of the electrons between the two
layers, which gives rise to an explicit asymmetry of the torque with
respect to current direction.  This asymmetry is an important
signature which can be used to distinguish spin-transfer-induced
torques from the torques produced by current-generated magnetic fields.
To see how the asymmetry arises, consider the
ferromagnet--normal-metal--ferromagnet (FNF) junction shown in Figure
\ref{figure2}. It consists of two ferromagnetic layers, F$_a$ and
F$_b$, with moments pointing in directions $\hat m_a$ and $\hat m_b$,
separated by a normal metal spacer N.  Normal metal leads on either
side of the trilayer inject an initially unpolarized current into the
system.  When the current enters the sample from the left (Fig.\
\ref{figure2}a), electrons transmitted through F$_a$ will be polarized
along $\hat m_a$.  As long as the normal metal spacer is smaller than
the spin-diffusion length (100 nm for Cu), this current will remain
spin-polarized when it impinges on F$_b$ and will exert a torque on
the moment of F$_b$ in a direction so as to align $\hat m_b$ with
$\hat m_a$. Repeating the argument for F$_b$, we find that the spin of
the electrons reflected from layer F$_b$ is aligned {\em antiparallel}
to $\hat m_b$, and, hence in turn, exerts a torque on the moment of
F$_a$ trying to align $\hat m_a$ antiparallel with $\hat
m_b$. (Subsequent multiple reflections of electrons between F$_a$ and
F$_b$ can serve to reduce the magnitudes of the initial torques, but
they do not eliminate or reverse them, as the electron flux is reduced
upon each reflection.)  When the current is injected from the right,
the directions of the torques are reversed: Now the flow of electrons
exerts a torque on F$_a$ trying to align its moment parallel with
$\hat m_b$, while it exerts a torque on F$_b$ so as to force the
moment in layer F$_b$ antiparallel with $\hat m_a$.

In the remainder of the paper, we assume that $\hat m_b$ points in the
$+z$ direction, while $\hat m_a$ differs by a small angle $\theta$ in
the $x - z$ plane.  (For thin films, demagnetizing forces will in
general cause the $y-z$ plane to be preferred, but this produces no
change in our argument.  We present the case that is easier to draw.)
The overall effect of a left-going flow of electrons then, is to exert
a torque $\vec \tau_b$ on F$_b$ in the $+x$ direction.  If we reverse
the current, so that electrons pass through F$_b$ first (Fig. 2b), the
torque on F$_b$ is only exerted by the electrons after they have been
reflected from F$_a$.  As seen before, the electrons reflected from
F$_a$ have polarizations opposite to $\hat m_a$, so that the torque on
F$_b$ is in the $-x$ direction.

In Refs.\ \onlinecite{slon1} and \onlinecite{myers}, the layer F$_a$
was taken to be much thicker than F$_b$, so that intralayer exchange
and anisotropy forces will hold the orientation of $\hat m_a$
fixed. In that case, one is only interested in the torque on F$_b$,
which serves to align $\hat m_b$ either parallel or antiparallel with the
fixed moment $\hat m_a$ depending on the current direction. This
asymmetric current response has been employed in both a point-contact
geometry\cite{myers} and in a thin-film pillar geometry
\cite{katine} to switch the moments in FNF trilayers from a parallel
to an antiparallel configuration by a current pulse in one direction,
and then from antiparallel to parallel by a reversed current.  For
weakly-interacting layers, either orientation can be stable in the
absence of an applied current, so that the resistance versus current
characteristic is hysteretic, and the devices can function as simple
current-controlled memory elements.

	
\section{Spin flux and torque in the scattering approach}
\label{2}

\begin{figure}[tbh]
\centerline{\psfig{figure=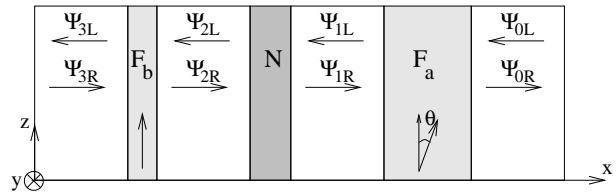,width=8.0cm}}
\vspace{3mm} \narrowtext
\caption[fig1]{\label{system} Schematic of the setup used for the
definition of the scattering matrices of the  F and N layers. The two
layers F$_a$
and F$_b$ are ferromagnetic layers whose magnetic moment is oriented as
shown in the bottom of the figure. The layer N is a nonmagnetic
metal spacer. Amplitude of left and right moving propagating waves
are defined in fictitious ideal leads
$0$, $1$, $2$ and $3$  between the  layers and between the layers and
the reservoirs.}
\end{figure}

	Treating the ferromagnetic layers as perfect spin filters provides
important qualitative insights into spin-transfer, but for a complete
qualitative and quantitative picture, a more general approach is required.
In this section, we introduce a scattering matrix description
of the FNF junction which allows us to deal with non-ideal
(magnetic and non-magnetic) layers. Our goal is to relate
 the torque $\vec{\tau}_b$ exerted on layer
F$_b$ by an unpolarized incident electron beam to the scattering properties
of the layers. Although we shall restrict our formulas to the FNF junction
 (see Fig.~\ref{system}), our method is applicable
for an arbitrary array of magnetic-nonmagnetic layers.

We first introduce the spin flux $\vec{J}$ in the $x$-direction
(the direction of current flow):
\begin{equation}
\label{spincurrent}
\vec{J}(x)=\frac{\hbar^2}{2m} {\rm Im} \int dy dz \left[ \phi^{\dagger}(x)
\vec{\sigma} \frac{\partial}{\partial x} \phi(x) \right]
\end{equation}
where $\phi (x)$ is a spinor wavefunction and $\vec{\sigma}$ the
vector of Pauli matrices,
$$\phi(x)=\left(\begin{array}{c} \phi_{\uparrow}(x)  \\
\phi_{\downarrow}(x) \end{array}
\right) ,\  \vec{\sigma}=\left( \begin{array}{c}
\sigma_x \\  \sigma_y \\ \sigma_z \end{array}
\right).
$$
Note that although Eq.(\ref{spincurrent}) bears close formal
analogy to the particle current, no local equation
of conservation can be written for the spin flux, since in
general, the Hamiltonian does not conserve spin. Specifically,
the magnetic layers can act as sources and
sinks of spin flux, so that the spin flux on different sides of
a F layer can be
different.

\subsection{Definition of the scattering matrices}

	Fig.~\ref{system} shows the FNF junction where (fictitious)
	perfect leads (labeled $0$, $1$, $2$ and $3$) have been added
        in between the layers F and N and between the F layers and
        the electron reservoirs on either side of the sample.
        The introduction of these leads allows for a description
	of the system using scattering matrices. In the perfect leads,
	the transverse degree of freedom are quantized, giving $N$
	propagating modes at the Fermi level, where $N \sim A/\lambda_F^2$
$A$ being the cross section area of the junction and $\lambda_F$ the
	Fermi wave length. Expanding the electronic wave function in these
	modes, we can describe the system in terms of
	the projection $\Psi_{i,L/R}$ of the wave function onto the
	left (right) going modes in region $i$. The $\Psi_{i,L/R}$
 	are $2N$-component vectors, counting the $N$ transverse modes and spin.
	The amplitudes of the wave function in two neighboring ideal
       leads are connected
	through the  scattering matrices $S_b$, $S_a$ and $S_{\rm N}$,
	that relate amplitudes of outgoing modes and incoming
	modes at the layer (see for
	example Ref. \onlinecite{beenakker1} for a review of
        the scattering matrix
	approach),

\begin{mathletters}
\label{defS}
\begin{eqnarray}
\left(
\begin{array}{c} \Psi_{3L} \\ \Psi_{2R} \end{array} \right)
& = & S_b
\left(\begin{array}{c} \Psi_{3R} \\ \Psi_{2L} \end{array}\right),  \\
\left(
\begin{array}{c} \Psi_{1L} \\ \Psi_{0R} \end{array}
\right) & = & S_a \left(
\begin{array}{c} \Psi_{1R} \\ \Psi_{0L} \end{array}
\right)  , \\
\left(
\begin{array}{c} \Psi_{2L} \\ \Psi_{1R} \end{array}
\right) & = & S_{\rm N} \left(
\begin{array}{c} \Psi_{2R} \\ \Psi_{1L} \end{array}
\right).
\end{eqnarray}
\end{mathletters}
The scattering matrices $S_b$, $S_a$ and $S_{\rm N}$ are $4N\times 4N$
unitary matrices. We decompose $S_b$ into $2N\times 2N$ reflection and
transmission matrices,
\begin{equation}
 S_b=\left(\begin{array}{cc} r_b & t'_b \\ t_b & r'_b \end{array}
\right),
\end{equation}
with similar decompositions of $S_a$ and $S_{\rm N}$.
 Normalization is done in such a way
that each mode carries unit current.  Due to the spin degree of
freedom, the reflection  and transmission
matrices can be written in terms of
four $N\times N$ blocks:
\begin{equation}
\label{rmatrix}
r_b= \left(
\begin{array}{cc} r_{b \uparrow \uparrow} & r_{b \uparrow \downarrow} \\
r_{b \downarrow \uparrow} & r_{b \downarrow \downarrow}  \end{array}
\right),
\end{equation}
where the subscripts $\uparrow, \downarrow$ refer to spin up and down in
the $z$-axis basis.

The scattering matrix of the magnetic layers depends
on the angle $\theta$ the moments may make with the z-axis.
The matrix $S_a(\theta)$ is related to $S_a(\theta=0)$ through a
rotation in spin space:
\begin{eqnarray}
r_a(\theta)& = &R_{\theta}\ r_a(0)\ R_{-\theta}, \ \
r'_a(\theta) =  R_{\theta}\ r'_a(0)\ R_{-\theta}, \nonumber \\
t_a(\theta)& = &R_{\theta}\ t_a(0)\ R_{-\theta}, \ \ 
t'_a(\theta) = R_{\theta}\ t'_a(0)\ R_{-\theta}
\end{eqnarray}
where
\begin{equation}
R_{\theta}= \left(
\begin{array}{cc} \cos \frac{\theta}{2} & -\sin \frac{\theta}{2}  \\
                  \sin \frac{\theta}{2} & \cos \frac{\theta}{2}
\end{array} \right) \otimes 1_N.
\end{equation}
The non-magnetic  metallic layer will not affect the spin states, {\em i.e.},
 $r_{\rm N\uparrow\downarrow}=r_{\rm N\downarrow \uparrow}=0$ and
$r_{\rm N \uparrow \uparrow}=r_{\rm N \downarrow \downarrow}$.

	We need to
keep track of the amplitudes within the system in order to calculate the
net spin flux deposited into each magnetic layer. Therefore, we define
$2N\times 2N$ matrices	$\Gamma_i^{L/R}$ and $\Lambda_i^{L/R}$ ($i=0,1,2,3$)
so that we may express
	all the $\Psi_{i,L/R}$ as a function of the amplitudes incident from the
two electrodes (regions 0 and 3):
\begin{equation}
\label{houlala}
\left(
\begin{array}{c} \Psi_{iL} \\ \Psi_{iR} \end{array} \right)
 =
\left(
\begin{array}{cc} \Gamma_{iL} & \Lambda_{iL} \\
                  \Gamma_{iR} & \Lambda_{iR} \end{array} \right)
\left(\begin{array}{c} \Psi_{0L} \\ \Psi_{3R} \end{array}\right)
\end{equation}
with the convention that $\Gamma_{0L}=\Lambda_{3R}=1$ and
$\Gamma_{3R}=\Lambda_{0L}=0$.
	In order to calculate the torque exercised on layer F$_b$ for
	a current entering from the left, we need
	the matrix, $\Gamma_{2L}$. To simplify the notations in the
	rest of the paper, we write
\begin{equation}
\Omega\equiv\Gamma_{2L}.
\end{equation}
 	The  matrix $\Omega$ relates $\Psi_{2L}$ to the incoming amplitudes
$\Psi_{0L}$ 	coming from the right. To calculate it, we put  $\Psi_{3R}=0$,
	then, using Eq.(\ref{defS}),  we get
        the  equations:
\begin{eqnarray}
\Psi_{1L} & = & t'_{a} \Psi_{0L} + r_{a} \Psi_{1R}, \nonumber \\
\Psi_{1R} & = & t_{n} \Psi_{2R} + r'_{n} \Psi_{1L}, \nonumber \\
\Psi_{2L} & = & r_{n} \Psi_{2R} + t'_{n} \Psi_{1L}, \nonumber \\
\Psi_{2R} & = & r'_{b} \Psi_{2L},  \nonumber \\
\Psi_{3L} & = & t'_{b} \Psi_{2L},
\end{eqnarray}
from which we obtain:
\begin{equation}
\label{omega}
\Omega=\frac{1}{1-r_nr'_b}t'_n
\frac{1}{1-r_at_nr'_b\frac{1}{1-r_nr'_b}t'_n -r_a r'_n}t'_a.
\end{equation}

\subsection{Spin flux response}

\begin{figure}[tbh]
\centerline{\psfig{figure=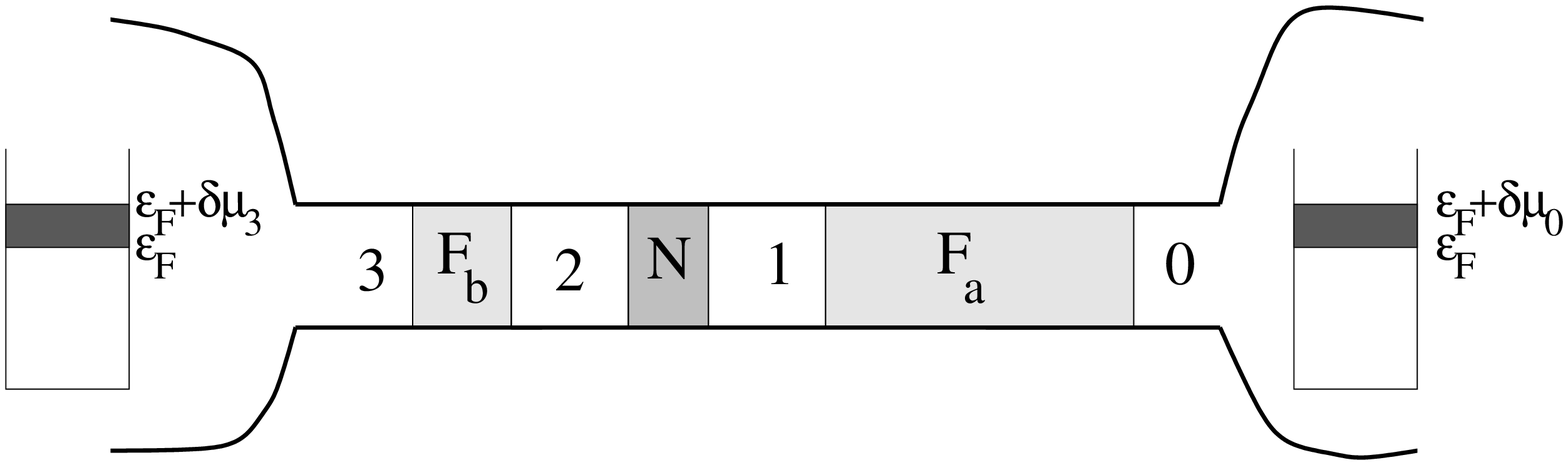,width=8.0cm}}
\vspace{3mm} \narrowtext
\caption[fig1]{\label{system2} The FNF junction  is connected
to two reservoirs.}
\end{figure}

	Let us now connect our system to two unpolarized electron reservoirs on its
two sides, as shown in Fig.~\ref{system2}. In equilibrium, the modes in
the reservoirs
are filled up to the fermi level $\epsilon_F$. We want to calculate
the spin current that is generated when the chemical potential in the
left (right) reservoir is slightly increased by $\delta\mu_3$
($\delta\mu_0$). The spin current $\vec{J_i}$ is the difference of the left
going
and right going contributions. For each of the region $i=0,1,2,3$, we
find from Eq.(\ref{spincurrent}) and Eq.(\ref{houlala}):
\begin{equation}
\label{aaa}
\frac{\partial\vec{J_i}}{\partial\mu_3}=\frac{1}{4\pi}{\rm Re }\left[ {\rm Tr}
\vec{\sigma} \Gamma_{iR}\Gamma_{iR}^{\dagger} -{\rm Tr} \vec{\sigma}
\Gamma_{iL}\Gamma_{iL}^{\dagger}\right],
\end{equation}
and
\begin{equation}
\label{bbb}
\frac{\partial\vec{J_i}}{\partial\mu_0}=\frac{1}{4\pi}{\rm Re }\left[ {\rm Tr}
\vec{\sigma} \Lambda_{iR}\Lambda_{iR}^{\dagger} -{\rm Tr} \vec{\sigma}
\Lambda_{iL}\Lambda_{iL}^{\dagger}\right].
\end{equation}
Derivation of Eq.(\ref{aaa}) and Eq.(\ref{bbb}) proceeds analogously to
the derivation of the Landauer formula for the conductance.\cite{BC}

\subsection{Torque exercised on layer F$_b$}

If the spin flux on both sides of F$_b$ (region 2 and 3) is different,
then angular momentum has been
deposited in the layer F$_b$. This creates a torque $\vec{\tau_b}$ on
the moment of the ferromagnet,
\begin{equation}
\vec{\tau_b} =  \vec{J}_3-\vec{J}_{2}.
\end{equation}

Setting $\delta\mu_0= -e V_0$,  we have:
\begin{equation}
\label{torque}
\frac{\partial\vec{\tau}_b}{\partial V_0}=-\frac{e}{4\pi} {\rm Re \ \
Tr}_{2N} \left[\vec{\Sigma} \Omega\Omega^{\dagger}\right],
\end{equation}
with
\begin{equation}
 \vec{\Sigma}=\vec{\sigma} -t'^{'\dagger}_b\vec{\sigma}t'_b
- r'^{\dagger}_b\vec{\sigma}r'_b.
\end{equation}
This equation can be simplified further if the spin-transfer effect is
due entirely
to spin-filtering (as argued by Slonczewski\cite{slon1}) as opposed
to spin-flip scattering of electrons from the
magnetic layers.  That is, if we assume that
$r_{b \uparrow \downarrow}=r_{b \downarrow \uparrow}=r_{a \uparrow
\downarrow}(\theta=0)=r_{a \downarrow \uparrow}(\theta=0)=0$,
then:
$$
\frac{\partial\tau_b^x}{\partial V_0} =-\frac{e}{2\pi} {\rm Re \ \ Tr}_{N}
\left
[
(\Omega_{\uparrow\uparrow}\Omega_{\downarrow\uparrow}^{\dagger}+\Omega_{\uparrow
\downarrow}
\Omega_{\downarrow\downarrow}^{\dagger})\right.
$$
\begin{equation}
\label{taux}
\left.
(1-r'_{b\uparrow\uparrow}
r'^{\dagger}_{b\downarrow\downarrow}
-t'_{b\uparrow\uparrow} t'^{\dagger}_{b\downarrow\downarrow} )\right]
\end{equation}
We will comment briefly in the conclusion of this paper about the physical
implications of including the off-diagonal spin-flip terms, as well.

We note that, as there is no spin flux
conservation in this system, $\partial\vec{J_i}/\partial\mu_3$
can be different from $- \partial\vec{J_i}/\partial\mu_0$ and,
 hence, there can be a non zero spin flux even when the chemical
potentials are identical in the two reservoirs.  The existence of a
zero-bias spin-flux and the resulting torques
reflect the well-known itinerant-electron-mediated exchange interaction
({\it a.k.a.}\ the RKKY interaction) between two ferromagnetic films
separated by a normal-metal spacer.  This interaction can in fact be
understood within a scattering framework.\cite{hathaway,slon93,bruno,edwards}
The zero-bias torque has to be added to the finite-bias contribution
(given by Eq.(\ref{taux})). Since the former is typically a factor
$N^{-1}$ smaller and vanishes upon ensemble averaging (see section
\ref{3} and Ref.\onlinecite{beenakker1}), we henceforth neglect the
zero-bias contribution to the
torque and restrict our attention to the bias induced torque, for
which we have
$$
\frac{\partial\vec{\tau}_b}{\partial V_0}
= -\frac{\partial\vec{\tau}_b}{\partial V_3}
$$
up to a correction of order $N^{-1}$.

\section{Averaging over the normal layer}
\label{3}

	Via Eq.(\ref{taux}), the torque on the moments of the
	ferromagnetic layers F$_a$ and F$_b$ not only depends on the
	scattering matrices $S_a$ and $S_b$ of these layers, but also
	of the scattering matrix $S_{\rm N}$ of the normal metal layer
	in between. If the normal layer is disordered, $\vec{\tau}_a$
	and $\vec{\tau}_b$ depend on the location of the impurities;
	if N is ballistic the torque depend sensitively on the
	electronic phase shift accumulated in N. In general, sample to
	sample fluctuations of the torque will be a factor $N^{-1}$
	smaller than the average.\cite{beenakker1} Hence, if $N$ is
	large ($N>10^3$ in the experiments of
	Ref.\onlinecite{myers}), the torque is well characterized by
	its average. In this section, we average over $S_{\rm N}$ for
	the case where N is disordered. The case of ballistic N is
	addressed in appendix B. After  averaging, the
zero-bias spin transfer current, corresponding to the RKKY interaction
described above, vanishes, and only the torque caused by the  electron
	current remains. Because all effects of quantum interference
	in the N layer will disappear in the process of averaging (to
	leading order in the number of modes $N$), the results we
	derive are unchanged if the reflection and transmission
	matrices include processes in which the energy of the electron 
        changes during  scattering,\cite{brataas} 
        in addition to the elastic proceses
	normally considered in scattering matrix calculations.

\subsection{Averaged torque}

The scattering matrix of the normal layer can be written using
the standard polar decomposition:\cite{StoneReview}
\begin{equation}
\label{polar}
S_n=
\left(
\begin{array}{cc} U & 0 \\ 0 & V'  \end{array}
\right)
\left(
\begin{array}{cc} \sqrt{1-T} & i\sqrt{T} \\ i\sqrt{T} & \sqrt{1-T}
\end{array}
\right)
\left(
\begin{array}{cc} U' & 0 \\ 0 & V  \end{array}
\right)
\end{equation}
where $U,V,U'$ and $V'$ are $2N\times 2N$ unitary matrices and T is a
diagonal matrix containing the eigenvalues of $t_nt_n^{\dagger}$.
Since $S_{\rm N}$ is diagonal in spin space, we find that $U$, $U'$,
$V$ and $V'$ are block diagonal.
\begin{equation}
\label{U}
U=
\left(
\begin{array}{cc} u & 0 \\ 0 & u  \end{array}
\right) ,
\ \
U'=
\left(
\begin{array}{cc} u' & 0 \\ 0 & u'  \end{array}
\right) ,
\end{equation}
and similar definitions for $v$ and $v'$. In the isotropic
approximation\cite{beenakker1,StoneReview}, the $N\times N$
unitary matrices $u, u', v$ and $v'$ are
uniformly distributed in the group $U(N)$.
(The outer matrices in Eq.(\ref{polar}) thus mix the modes
in a ergodic way while the
central matrix contains the transmission properties of the layer,
which determine  the average conductance of
N.)

  We want to average Eq.(\ref{taux}) over both the unitary matrices
and $T$. A diagrammatic technique for such averages has already been
  developed in Ref.\onlinecite{brouwer1} and can be used to
  calculate $\langle\partial\vec{\tau}_b/\partial V_0\rangle$ in leading
  order in $1/N$. It is a general property of such averages that the
fluctuations are a factor of order $N$ smaller than the average. This
justifies our statement above, that the ensemble averaged torque is
sufficient to characterize the torque exerted on a single sample.
Details of the calculation are presented in Appendix A.

  The resulting expression
  for  $\langle\partial\vec{\tau}_b/\partial V_0\rangle$ can be
written in a form very similar to the one for
  Eq.(\ref{taux}) if one uses a notation that involves $4\times 4$
  matrices. To be specific, to each $2N\times 2N$ matrix $A$
  appearing in Eq.(\ref{taux}) and Eq.(\ref{omega}), we assign a
  $4\times 4$ matrix $\hat{A}$ as,
\begin{equation}
\label{coucou}
\hat{A}=\frac{1}{N} {\rm
Tr}_{N} \left[A \otimes A^{\dagger}
\right],
\end{equation}
where ${\rm Tr}_N$ means that the trace has been taken in each the
$N\times N$ blocks, or in extenso:
\begin{equation}
\hat{A}=\frac{1}{N} {\rm Tr}_N
\left(
\begin{array}{cccc} A_{\uparrow \uparrow}A_{\uparrow \uparrow}^{\dagger} &
A_{\uparrow \uparrow}A_{\uparrow \downarrow}^{\dagger} &
 		    A_{\uparrow \downarrow}A_{\uparrow \uparrow}^{\dagger} &
A_{\uparrow \downarrow}A_{\uparrow \downarrow}^{\dagger} \\
 		    A_{\uparrow \uparrow}A_{\downarrow \uparrow}^{\dagger} &
A_{\uparrow \uparrow}A_{\downarrow \downarrow}^{\dagger} &
 		    A_{\uparrow \downarrow}A_{\downarrow \uparrow}^{\dagger} &
A_{\uparrow \downarrow}A_{\downarrow \downarrow}^{\dagger} \\
                    A_{\downarrow \uparrow}A_{\uparrow \uparrow}^{\dagger}
&  A_{\downarrow \uparrow}A_{\uparrow \downarrow}^{\dagger} &
 		    A_{\downarrow \downarrow}A_{\uparrow \uparrow}^{\dagger} &
A_{\downarrow \downarrow}A_{\uparrow \downarrow}^{\dagger} \\
                    A_{\downarrow \uparrow}A_{\downarrow
\uparrow}^{\dagger} &  A_{\downarrow \uparrow}A_{\downarrow
\downarrow}^{\dagger} &
 		    A_{\downarrow \downarrow}A_{\downarrow \uparrow}^{\dagger} &
A_{\downarrow \downarrow}A_{\downarrow \downarrow}^{\dagger}
\end{array} \right).
\end{equation}
We also define $\hat{\vec{\Sigma}}$ by,
\begin{equation}
\hat{\vec{\Sigma}}= {\rm Tr}_N
\left(
\begin{array}{cccc}
\vec{\Sigma}_{\uparrow \uparrow}    & \vec{\Sigma}_{\downarrow \uparrow}  &
\vec{\Sigma}_{\uparrow \downarrow}  & \vec{\Sigma}_{\downarrow \downarrow}  \\
 		        0 & 0  & 0 & 0 \\
  			0 & 0  & 0 & 0 \\
\vec{\Sigma}_{\uparrow \uparrow}    & \vec{\Sigma}_{\downarrow \uparrow}  &
\vec{\Sigma}_{\uparrow \downarrow}  & \vec{\Sigma}_{\downarrow \downarrow}
\\
\end{array} \right).
\end{equation}
The average over the transmission eigenvalues $T$ follows if we note
that
the average of a function is the function of the average, to leading
order in $1/N$.\cite{beenakker1} Thus the average over $T$ amounts to the
replacement
\begin{equation}
\hat{t}_n= \frac{g_{\rm N}}{N} \openone_4 {\rm \ \ and \ \ } \hat{r}_n=
 \left(1-\frac{g_{\rm N}}{N}
\right) \openone_4,
\end{equation}
where $g_{\rm N}$ is the conductance of the normal layer and
$\openone_4$ is the $4\times 4$ unit matrix.
Using these ``hat'' matrices, the result has now the simple form:
\begin{equation}
\label{main}
 \langle\frac{\partial\vec{\tau}_b}{\partial V_0}\rangle
=-\frac{e}{4\pi} {\rm Re \ \
Tr}_{4} \left[\hat{\vec{\Sigma}} \hat{\Omega}\right],
\end{equation}
where (compare to Eq.(\ref{omega})),
\begin{equation}
\hat{\Omega}=\frac{1}{1-\hat{r}_n\hat{r}'_b}\hat{t}'_n
\frac{1}{1-\hat{r}_a\hat{t}_n\hat{r}'_b\frac{1}{1-\hat{r}_n\hat{r}'_b}\hat{t
}'_n -\hat{r}_a \hat{r}'_n}\hat{t}'_a.
\end{equation}

Equation (\ref{main}) is the main result of this paper.
In the absence of spin-flip scattering, it reduces to

\begin{eqnarray}
\langle\frac{\partial\tau_b^x}{\partial V_0}\rangle
& = &-\frac{e}{2\pi} {\rm Re }
\left[ (\hat{\Omega}_{3,1}+\hat{\Omega}_{3,4})\right. \nonumber \\
& \times &\left.{\rm Tr}_{N}(1-r'_{b \uparrow \uparrow} r'^{\dagger}_{b \downarrow
\downarrow}
-t'_{b \uparrow \uparrow} t'^{\dagger}_{b \downarrow \downarrow} )\right].
\end{eqnarray}
The same formalism can be used to calculate the conductance $g$ of the
system using the Landauer formula. One gets:
\begin{equation}
\label{defg}
\langle g \rangle= \frac{N e^2}{h} \left[\hat{t'}_{1,1}+\hat{t'}_{1,4}
+\hat{t'}_{4,1}+\hat{t'}_{4,4}\right],
\end{equation}
$t'$ being the total transmission matrix:
\begin{equation}
t'=t'_b \ \Omega .
\end{equation}

We would like to note that, while our theory started from a fully
phase coherent description of the FNF trilayer, including the full
$4N\times 4N$ scattering matrices of the FN interfaces, the final
result can be formulated in term of $2\times 4$ parameters, 
represented by the matrices $\hat{r}_a$ and $\hat{r}'_b$ 
( $2\times 16$ parameters in case of
spin-flip scattering). Such a reduction of the number of degrees of
freedom was also found by Brataas {\it et
al.},\cite{brataas} although their starting point is an 
hybrid ferromagnetic-normal metal circuit  with incoherent
nodes. This confirms the statement at the beginning of this section,
that for a diffusive normal-metal spacer all effects of quantum
interferences are washed out.\cite{beenakker1} The difference between
our approach and the one of Ref.\ \onlinecite{brataas} is important in
the case of the ballistic normal layer, see Appendix B.

\subsection{Symmetries}
Before we proceed with a further analysis of Eq.(\ref{main}), we
identify the different symmetries of the torque. Due to
the conservation of current, the total torque deposited on the full system
is anti-symmetric with respect to current direction:
\begin{equation}
\label{somme}
\frac{\partial\vec{\tau}_b}{\partial
V_0}+\frac{\partial\vec{\tau}_a}{\partial V_0}
=
-\left[\frac{\partial\vec{\tau}_b}{\partial V_3}+
\frac{\partial\vec{\tau}_a}{\partial V_3}\right].
\end{equation}
Eq.(\ref{somme}) holds before averaging. But, as
pointed out in section \ref{2}, equality for each of the
torques $\vec{\tau}_a$ and $\vec{\tau}_b$ separately holds
only after averaging,
\begin{equation}
\langle\frac{\partial\vec{\tau}_b}{\partial V_0}\rangle
=-\langle\frac{\partial\vec{\tau}_b}{\partial V_3}\rangle;
\end{equation}
sample to sample fluctuations of $\partial\vec{\tau}_b/\partial V_0$
and $-\partial\vec{\tau}_b/\partial V_3$ of relative order $1/N$ are
in general different. Thus, for $N\gg 1$, our calculation can be used
to compute the linear response of the torque to a small bias voltage:
\begin{equation}
\vec{\tau}_b = \langle\frac{\partial\vec{\tau}_b}{\partial V_0}\rangle
( V_0 - V_3).
\end{equation}
In our geometry, where F$_a$ and F$_b$ are in $x-z$ plane,
the only non-zero component of the torque is $\tau_b^x$.
The torque vanishes when the moments are completely aligned or
antialigned (all the matrices are diagonal in spin space and therefore
no x-component of the spin can be found). Around these two limits, the
torque is symmetric in respect to the angle ($\theta \rightarrow
-\theta$ and $\pi-\theta \rightarrow \pi +\theta$).
There is no symmetry between $\theta$ and $\pi -\theta$. In
addition, the two layers are not equivalent and exchanging the
scattering matrices of F$_a$ and F$_b$ also changes the torque.

\subsection{Discussion of some limiting cases}
Eq.(\ref{main}) can be simplified in some particular
cases. Let us start with the simplest case of ideal spin filters, so that
majority (minority) spins are totally transmitted (reflected) by either layer.
Equation (\ref{main}) then reduces to
\begin{equation}
\label{clean}
\langle\frac{\partial\tau_b^x}{\partial V_0}\rangle  =
- \frac{e}{4\pi} \frac{g_{\rm N} \sin\theta}{3+\cos\theta}=
-\frac{h}{4\pi e} \langle g \rangle \frac{\tan \theta /2}{2},
\end{equation}
where $\langle g \rangle$ is the average magnetoconductance,
cf Eq.(\ref{defg}),
\begin{equation}
\langle g \rangle  =  \frac{e^2}{h} g_{\rm
N}\frac{4\cos^2\theta/2}{3+\cos\theta}.
\end{equation}
Equation~(\ref{clean})  reproduces a result of Slonczewski.\cite{slon1}
As expected, for left-going
electrons ($V_0<0$)
the torque is positive, so it acts to align the moments of the two magnetic
layers, see section~\ref{1}.

	Let us now consider the case of weak $s-d$ exchange coupling, i.e.,
when the scattering coefficients depend only weakly on spin. We continue to
assume
that no spin-flip
scattering occurs in the ferromagnetic layers. We define $g_a$ and $g_b$
as the average conductance  per spin of the two layers
(in unit of $e^2/h$). Then, the  conductance of F$_a$ alone is
$g_a+\delta g_a$ and $g_a- \delta g_a$ for respectively the majority
and minority spins, which defines the spin scattering asymmetry $\delta g_a$.
In that case, we get to lowest nontrivial order in $\delta g_a$
and $\delta g_b$:
\begin{eqnarray}
\langle g\rangle & = &\frac{2e^2}{h}\left(\frac{g_{\rm N} g_a g_b}
{g_a g_b + g_{\rm N}(g_a+g_b-2\frac{g_a g_b}{N})}\right. \nonumber \\
 & + & \left. \frac{g_{\rm N}^2 \delta g_a
\delta g_b \cos \theta}{(\frac{g_a g_b}{N} + g_{\rm N}(g_a+
g_b-2\frac{g_a g_b}{N}))^3}\right)
\end{eqnarray}
and
\begin{equation}
\langle\frac{\partial\tau_b^x}{\partial V_0}\rangle=
- \frac{e}{2\pi}\frac{g_{\rm N}^2 \delta g_a \ \delta g_b^2 \sin\theta}
{2(1-\frac{g_b}{N})(g_a g_b + g_{\rm N}(g_a+g_b-2\frac{g_a g_b}{N}))^2}.
\end{equation}
This last formula shows that:

(i) The torque is not symmetric with respect to interchanging the
layer F$_a$ and
F$_b$, in contrast to the conductance. If one changes $\delta g_a$ to
$-\delta g_a$, the sign of the torque is reversed.  However,
$\langle\partial\tau_b^x/\partial V_0\rangle \propto\delta g_b^2$, so if
one changes $\delta g_b$ to
$-\delta g_b$, the sign of the torque is unchanged.  The sign of the torque
on a ferromagnetic layer therefore depends on whether the
other layer is a positive or negative
polarizer, but not on the sign of filtering for the layer experiencing the
torque. We have verified that this is true also in the general
case. This point explains why the two layers can not be treated on an equal
footing.

(ii) We see that
$g_{\rm N}$ appears through its square. Indeed, in order for some spin to be
deposited in the layer F$_b$, some left going electrons have to be
reflected by F$_b$ and exit the system from the right hand side.
Therefore these electrons cross the normal layer at least twice and
this leads to the factor $g_{\rm N}^2$. On the other hand the conductance
is linear in $g_{\rm N}$. Therefore in order to maximize the
torque deposited per current,
one has to use the cleanest possible normal
metal spacer. (This statement is true in this limit of weak
filtering, but not in general, see
section \ref{5}.) Note that in the previous case (perfect
spin-filtering) the torque is proportional to  $g_{\rm N}$
instead of the expected $g_{\rm N}^2$. Indeed, in that case,
once the electron has been reflected by the layer F$_b$, it cannot
go through F$_b$ which works as a perfect wall for it. Therefore
current conservation implies that it goes out of the system through the
right. For $g_{\rm N}\ll N$, the torque is actually proportional to
$g_{\rm N}^2$ for arbitrary spin asymmetry (except perfect filtering),
and one gets:
\begin{equation}
\langle\frac{\partial\tau_b^x}{\partial V_0}\rangle \propto g_{\rm N}^2
\sin \theta, \ \ \ g_{\rm N} \ll N,
\end{equation}
the factor of proportionality being a complicated function of the
transmission probabilities of the layers.

\section{Application to current-driven switching of magnetic domains}
\label{5}
In this section, we consider the general solution Eq.(\ref{main})
for the spin-transfer torque. We first address  strongly polarizing
systems and
then calculate torques for scattering
parameters more appropriate for the transition metal trilayers
that can be studied experimentally. As the trilayer devices are primarily
current-driven,
we calculate the torque per unit of current $I$,
$$
\frac{\tau_b^x}{I}=
\langle \frac{1}{g} \frac{\partial\tau_x}{\partial V_0}\rangle.
$$
The torque is measured in units of $\frac{h}{2\pi e}$.
\begin{figure}[tbh]
\centerline{\psfig{figure=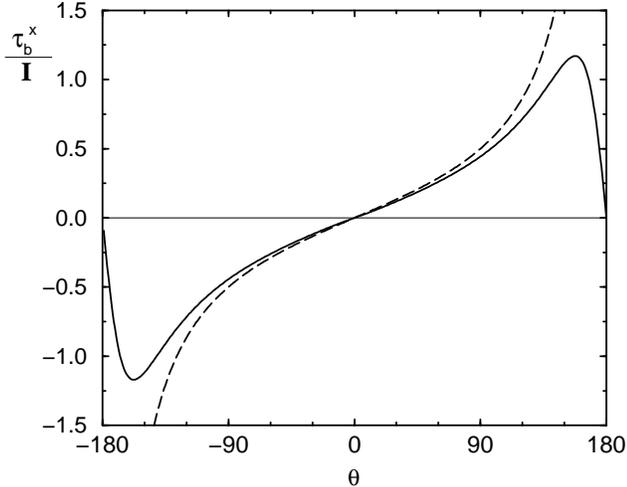,width=9.0cm}}
\vspace{3mm} \narrowtext
\caption[fig1]{\label{perfectb-t} Torque per unit current for the case
where F$_b$ is a nearly perfect polarizer ($|t_{b\uparrow}|^2=0.999$,
$|t_{b\downarrow}|^2=0.001$) and F$_a$ is not ($|t_{a\uparrow}|^2=0.3$
  $|t_{a\downarrow}|^2=0.01$). (solid line). The dashed line shows the
case of perfect polarizers, see Eq.(\ref{clean}).
Torque in measured in unit of $\frac{h}{2\pi e}$.}
\end{figure}

Eq.~(\ref{clean}) of the previous section gives the torque per unit
current for the case that both layers F$_a$ and F$_b$ are perfect
polarizers.
The main feature of this system is that
the $\theta$ dependence of the torque is not of a simple $\sin \theta$
form, and that the torque per unit current diverges at $\theta=0$.
In Fig.~\ref{perfectb-t}, we look at what happens when one of the layers
(F$_b$) is a nearly perfect polarizer while the other one
is not.  Although the divergence at $\theta=\pi$ is regularized,
 $\tau_b^x/I$ remains sharply peaked near $\theta=\pi$. This
is relevant for the critical current needed  to switch the magnetization of
F$_b$ from $\theta=\pi$ to $\theta=0$. Recall that the switching of
the domains follows from a competition between the spin-transfer
torque on the one hand and restoring forces from local fields,
anisotropy, exchange coupling etc. (The competition between these
forces has been considered  phenomenologically in
Ref.~\onlinecite{katine,slon2}
using a  phenomenological Landau-Lifschitz-Gilbert equation.)
The torques for $\theta$ close to $0$ and $\pi$ determine the
critical currents to overturn a metastable parallel (antiparallel)
alignment
of the moment in F$_a$ and F$_b$. Hence the critical current should
be different at $\theta=0$ and $\theta=\pi$.
\begin{figure}[tbh]
\centerline{\psfig{figure=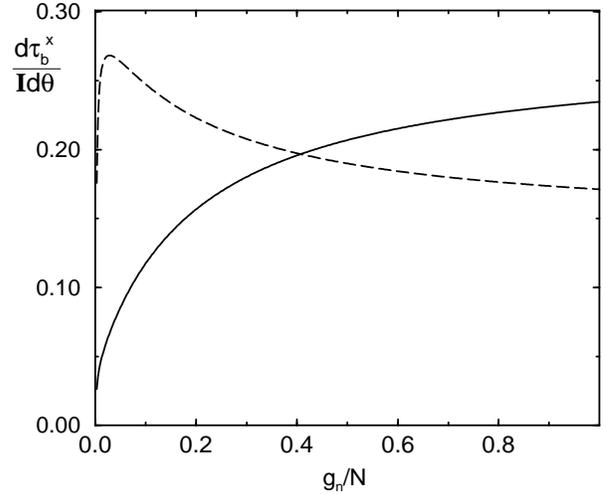,width=9.0cm}}
\vspace{3mm} \narrowtext
\caption[fig1]{\label{perfectb-gn} Derivative of the torque with
respect to $\theta$ at $\theta=0$ as a function of $g_{\rm N}$, in
unit of $\frac{h}{2\pi e}$ for the case where
F$_b$ is a nearly perfect polarizer ($|t_{b\uparrow}|^2=0.999$,
$|t_{b\downarrow}|^2=0.001$) and F$_a$ is not ($|t_{a\uparrow}|^2=0.3$
  $|t_{a\downarrow}|^2=0.01$) (solid line), and for the opposite setup,
F$_a$ is a nearly perfect polarizer ($|t_{a\uparrow}|^2=0.999$,
$|t_{a\downarrow}|^2=0.001$) while F$_b$ is not ($|t_{b\uparrow}|^2=0.3$
  $|t_{b\downarrow}|^2=0.01$) (dashed line).
}
\end{figure}

In Fig.~\ref{perfectb-gn}, we consider the same system as in
 Fig.~\ref{perfectb-t} (one perfectly polarizing F layer, one
partially polarizing layer), but as a function of the conductance of the
normal layer $g_{\rm N}$ for angles $\theta$ close to $0$.
We find that switching the two layers
has a drastic effect on the torque, even at a qualitative level.
Interestingly, in the case where F$_a$ is the
nearly perfect layer (dashed line in Fig.~\ref{perfectb-gn}),
a maximum of the torque is found
for  $\frac{g_{\rm N}}{N}\ll 1$, i.e., in that case, a dirty metal
 spacer would give a higher torque (per unit of current) than a
clean one.

\begin{figure}[tbh]
\centerline{\psfig{figure=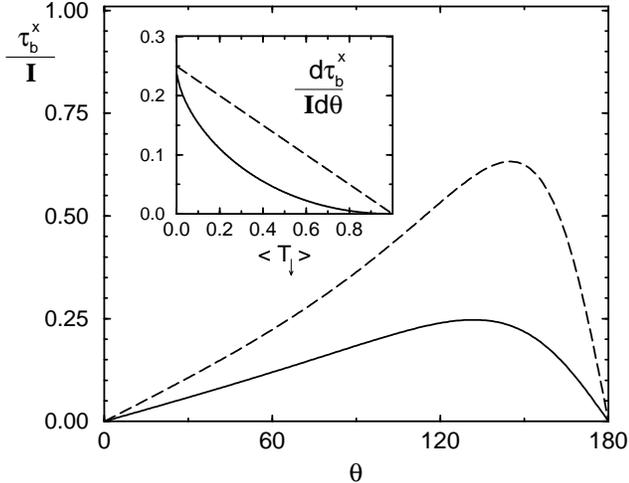,width=9.0cm}}
\vspace{3mm} \narrowtext
\caption[fig1]{\label{sloncz}
Torque per unit current as a function of $\theta$.
The solid line shows the case where
the minority spin are transmitted with probability $\langle
T_{\downarrow} \rangle=20\%$ for all the channels. The dashed line
shows the case where the minority spins are transmitted with
probability  $\langle
T_{\downarrow} \rangle=1$ for $20\%$ of the channels, $0$ otherwise
(theory of Ref.~\onlinecite{slon1}).
The majority spin are totally transmitted in both cases.
Inset: same system,
$d\tau_b^x/Id\theta$ at $\theta=0$ as a function of $\langle
T_{\downarrow} \rangle$ for the two different models.}
\end{figure}

At this stage, it is interesting to compare our
theory to that of Ref~\onlinecite{slon1}. In this work the
WKB approximation was used, and the electrons at the FN
interfaces are either totally transmitted or
reflected. For non-perfect polarizers, only a fraction
of the channels\cite{noteslon}
 act as
perfect filters while the others perfectly transmit both the minority and
majority spins. However, this situation is different from having non
perfect transmission probabilities $T_{\downarrow}$, $T_{\uparrow}$
per channel. In particular, having $\langle
T_{\downarrow} \rangle N$ channels that do not filter and $(1-\langle
T_{\downarrow} \rangle ) N$ perfect filters is not equivalent to $N$
channels that all partly transmit the minority spins with probability
$\langle T_{\downarrow} \rangle$. This situation
is illustrated in Fig~\ref{sloncz}. The latter scenario is supported
by {\it ab-initio} calculations.\cite{stiles,schep} Moreover, for a
disordered normal-metal spacer, multiple scattering from impurities
mixes all channels and the notion of two type of channels become
superfluous. In that case, the torque is described by Eq.(\ref{main})
in all cases. The torque found in the second
case can be significantly smaller than under the  assumption of
Ref~\onlinecite{slon1}.

We can also compare our model to the work of Berger.\cite{berger}  While
the theories of Berger and Slonczewsi\cite{slon1} have much in common,
Berger does invoke inelastic spin-flip scattering in a way that Slonczewski
does not.  (Slonczewski's theory utilizes only spin-filtering, without
spin-flip scattering.)  This effect can in principle be treated in our
model, by including the off-diagonal spin-flip reflection and transmission
amplitudes that we have thus far neglected.  We shall comment on some of
the implications in the conclusion.  We suspect that the differing
treatments of this aspect of the physics may explain why Slonczewski and
Berger predict slightly different forms for the current-induced torques. 

\begin{figure}[tbh]
\centerline{\psfig{figure=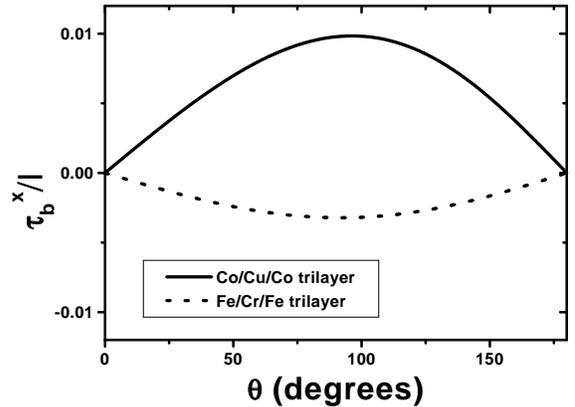,width=9.0cm}}
\vspace{3mm} \narrowtext
\caption[fig1]{\label{realistic}
Torque per unit current as a function of $\theta$ for two different
realistic systems. The solid line shows the Co-Cu-Co trilayer
($|t_{a\uparrow}|^2=0.73$,
$|t_{a\downarrow}|^2=0.49$, $|t_{b\uparrow}|^2=0.68$,
$|t_{b\downarrow}|^2=0.29$). The dashed line shows the Fe-Cr-Fe trilayer
($|t_{a\uparrow}|^2=0.48$,
$|t_{a\downarrow}|^2=0.59$, $|t_{b\uparrow}|^2=0.30$,
$|t_{b\downarrow}|^2=0.50$). In both cases $g_{\rm N}=N$ has been
assumed. The thick layer is assumed to be semi-infinite while for the
thin layer, only the interface properties have been taken into account.
Numerical values are obtained from Ref.~\onlinecite{stiles}.
}
\end{figure}

In our theory, the scattering matrices of the ferromagnetic layers
appear as free input parameters. However, it is in principle possible to
calculate them from first principle calculations for specific
materials. Such an approach has
been taken in Ref.~\onlinecite{stiles,schep} and the results
can be used to give some
estimates of  torques that can be expected in realistic systems.
In Fig.~\ref{realistic}, we compare  the Co-Cu-Co system considered in the
experiment of Ref. \onlinecite{myers} with the Fe-Cr-Fe system. In the latter,
the minority spins have a larger transmission probability than the
majority ones, explaining
the opposite sign of the torque.

\section{Conclusion}
\label{6}

We have developed a theory  for the spin-transfer-induced torques on
the magnetic moments of  a
ferromagnet-normal-ferromagnet FNF trilayer system caused by a flowing current.
Our theory deals with the effects of multiple scattering between
the layers using the scattering matrices of the ferromagnet-normal
metal interfaces as input parameters. We consider both the cases of a
diffusive and ballistic normal metal spacer. Remarkably, in the
diffusive case,  the high-dimensional scattering matrices of the
FN interfaces only appear through the reduced $4\times 4$ tensor products of
Eq.(\ref{coucou}) which greatly reduces the number of degrees of freedom
of the theory (see also Ref.\
\onlinecite{brataas}). This reduction of the number of degrees of freedom
allows us to make qualitative predictions about the role of the
interface transparency, normal metal resistance etc., without detailed
knowledge of the microscopic details of the system. However, for
quantitative predictions, inclusion of the microscopic parameters in
our theory, e.g. from {\it ab initio} calculations\cite{stiles,schep} is
still  needed.

	Having a complete theoretical description of the
current-induced 
switching of magnetic domains in FN multilayers as
a final goal, the theory here can be regarded as being an intermediate
	step. On the one hand, microscopic input is needed for the
scattering matrices of the FN interfaces, as explained above. On the
	other hand, the output of our theory, the current-induced
	torques, needs to be combined with restoring (hysteretic)
	forces in a more phenomenological theory that describes the
dynamics of the magnetic moments. Such a theory involves anisotropy
	forces and information about the mechanism by which the torque
	is exerted (spin wave excitation, local exchange field) --
	issues which are still subject of debate.\cite{berger,slon1,heide,bazaliy}.

 In this paper, we have focused on the effects of ``spin filtering'' as the
mechanism for current-induced torque, {\it i.e.}, the difference in the
transmission and reflection probabilities for electrons with spins
parallel and antiparallel to the moments of the ferromagnetic layers (the
diagonal terms in the matrices for the reflection and transmission
amplitudes, Eq.(\ref{rmatrix}).) A different source of
spin-dependent scattering, which we have not considered in detail, but
which is included in our formalism, is that of spin-flip
scattering -- the off-diagonal terms in Eq.(\ref{rmatrix}). Its effect can be twofold. In the normal spacer, it would
decrease the effective
polarization, and therefore the torque. However, in the ferromagnet,
the rate of spin-flip scattering might
be asymmetric with respect to minority and majority spins, and therefore
spin-flip scattering may also be an additional  source of torque.
As the number of degrees of freedom involved is
much larger than for spin filtering only, a realistic model for the
scattering matrices in the ferromagnets would be a necessary starting
point for a theory that would include the effect of spin-flip scattering.
We leave such a theory for future work.

\acknowledgments
We thank A. Brataas and G. Bauer for drawing our attention 
to Ref.~\onlinecite{brataas}.

\section*{APPENDIX}

\subsection{Derivation of Eq.(\ref{main})}

 In this appendix, we describe the calculation of Eq.(\ref{main})
step by step.

	First, we substitute the expression (\ref{omega})
for $\Omega$
into (\ref{torque}), and then formally expand the resulting equation
in powers of the reflection matrices $r_a$, $r_b$, $r_n$ and $r'_n$.
Using the polar decomposition Eq.(\ref{polar}) for the reflection and
transmission matrices  $r_n, t_n, t'_n$ and $r'_n$ of the normal
layer, we get a sum of many terms, each of which is of a form
where contributions from N are alternated with those of F$_a$ and
F$_b$. Writing spin indices explicitly (summation over repeated indices
is implied), we can write those terms as,
\begin{equation}
\label{step2}
{\rm Tr}_{N} \vec{\Sigma}_{ij} \left( A_{jk} \alpha
B_{kl}\beta... \eta C_{sm} \right)
\left(F^{\dagger}_{nm} \omega^{\dagger}...\delta^{\dagger} E_{pn}^{\dagger}
\gamma^{\dagger} D_{ip}^{\dagger}\right),
\end{equation}
where $A$, $B$, $C$, $D$, $E$, $F$ $\in \{r_a,t_a,r_b,r'_b...\}$ refer to
the layer
F$_a$ and F$_b$ while $\alpha$, $\beta$, $\gamma$, $\delta$, $\eta$,
$\omega \in
\{ui\sqrt{T}v , u\sqrt{1-T}u',...\}$ refer to the normal layer.

        We are now ready to do the average of Eq.(\ref{step2})
over the matrices $u, u', v$ and $v'$ using the diagrammatic technique
of Ref.~\onlinecite{brouwer1} (In leading order in $N$ these integrals
reduce to the application of wick theorem.) Doing so, each of the
	$\alpha,\beta,...$ has to be put in correspondence with one of
	the $\gamma^{\dagger}, \delta^{\dagger}$, etc. To leading order
	in $N$, only the ladder diagram survives, in which
	$\alpha=\gamma$, $\beta=\delta$, $\eta=\omega$,... and hence,
    $A=D$, $B=E$, $C=F$,... Thus, after averaging,  we get terms like:
\begin{eqnarray}
\label{step3}
{\rm Tr}_{N} \left[\vec{\Sigma}_{ij}\right]
\frac{1}{N} {\rm Tr}_{N}\left[A_{jk}A_{ip}^{\dagger}\right] a
\frac{1}{N} {\rm Tr}_{N}\left[B_{kl}B_{pn}^{\dagger}\right] b
\nonumber\\
...\ \ \ c \frac{1}{N} {\rm Tr}_{N}\left[C_{sm}C_{nm}^{\dagger}\right],
\end{eqnarray}
where $a,b,c,...$ stands for either $\frac{1}{N}{\rm Tr} T$  or
$\frac{1}{N}{\rm Tr} (1-T)$.
	
	To leading order in $N$,  the average over $T$ can now be done
by simply replacing
$a,b,c,...$ by their average value $g_{\rm N}/N$ or $1-g_{\rm N}/N$ where
$g_{\rm N}$ is the
average conductance (per spin) of the normal layer, in unit of
$e^2/h$.

	Finally, denoting $\lambda=(i,j)$ and $\mu=(k,p)$, let us now
introduce $4\times 4$ matrices $\hat{A}$, $\hat{B}$, $\hat{C}$,...
that are defined as:
\begin{equation}
\label{defhat}
\hat{A}_{\lambda\mu}=\frac{1}{N} {\rm
Tr}_{N}\left[A_{ik}A_{jp}^{\dagger}
\right],
\end{equation}
and $\hat{\vec{\Sigma}}$ is defined as
\begin{equation}
\hat{\vec{\Sigma}}_{\lambda\mu}= \delta_{kp}
{\rm Tr}_{N}\vec{\Sigma}_{ji}.
\end{equation}
In term of these new matrices, eq.(\ref{step3}) now reads as a simple matrix
product:
\begin{equation}
\label{step5}
{\rm Tr}_4\hat{\vec{\Sigma}}
\hat{A}\hat{\alpha}\hat{B}\hat{\beta}...\hat{\eta}\hat{C},
\end{equation}
with  $\hat{\alpha}$, $\hat{\beta}$, $\hat{\eta}$,...$\in
\{g_{\rm N}/N,1-g_{\rm N}/N\}$.
Equation (\ref{step5}) is formally equal
to the expansion of $\Omega$ (see Eq.(\ref{step2})) except
that we are now dealing with ``hat'' matrices. Therefore, we can now
resum all the terms of the expansion and get Eq.~(\ref{main}).

\subsection{Ballistic normal layer: a pedestrian approach}
\label{4}

If N is very clean, and the interfaces are very flat,
it is reasonable to assume that the electrons propagate ballistically
inside the normal layer. The different modes will not be mixed in that
case, and the electron wavefunction only picks up a phase factor
$e^{ik_iL}$ where $L$ is the width of N and $k_i$ the momentum
of channel $i$. For a sufficiently thick normal layer (i.e. $L\gg
\lambda_F$), small fluctuations of
$k_i$ lead to an arbitrary change in the phase factor, and it is
justified to consider $e^{ik_iL}$ as a random phase and to average over
it. This is different from the case of a disordered metal spacer,
where the
average involves unitary matrices $u$, $u'$... that mix the channels,
cf Eq.(\ref{polar}). In the case where $r_{a\downarrow}$,
$r_{a\uparrow}$,... are proportional to the identity matrix
(i.e. the reflection amplitudes do not depend on the channel), the
ballistic model reduces to the disordered model of Eq.~(\ref{main}) for
$g_{\rm N}=N$.

The reflection matrices of N being zero, the matrix $\Omega$  reads:
\begin{equation}
\Omega= e^{ik_iL}
\frac{1}{1-e^{2ik_iL} r_a r'_b}t'_a.
\end{equation}
Neglecting spin-flip scattering,  denoting $z=e^{2ik_iL}$,
and choosing $r_{a11}=r_{a\uparrow}$, $r_{a22}=r_{a\downarrow}$,... where
$r_{a\uparrow}$, $r_{a\downarrow}$,... are diagonal matrices, one gets
after some algebra:
\begin{equation}
\frac{\partial\tau_b^x}{\partial V_0}(z)=
- \frac{e\nu}{4\pi}{\rm Tr \ Re \ } \frac{ A(z)}{z |D(z)|^2}\sin\theta,
\end{equation}
where $A(z)$ and $D(z)$ stand for:
\begin{eqnarray}
A(z) &=&
\left( 1-t'_{b\uparrow}t'^*_{b\downarrow} -
r'_{b\uparrow}r'^*_{b\downarrow} \right) \nonumber \\
 &(& |t'_{a\uparrow}|^2
(1-z r'_{b\downarrow}r_{a\downarrow})
(z- r'^*_{b\uparrow} r_{a\downarrow}^*) \nonumber \\
  & - & \left. |t'_{a\downarrow}|^2(1-zr_{b\downarrow}r_{a\uparrow})(z-
r'^*_{b\uparrow}r_{a\uparrow}^*) \right), \\
D(z) & = &1-z \left[ \cos^2\frac{\theta}{2} (r_{a\uparrow}r'_{b\uparrow}
+ r_{a\downarrow} r'_{b\downarrow})\right. \nonumber \\
 & + & \left. \sin^2\frac{\theta}{2}
(r_{a\downarrow}r'_{b\uparrow}+r_{a\uparrow}r'_{b\downarrow})\right]
\nonumber \\
& + & z^2 r_{a\downarrow}r_{a\uparrow}r'_{b\downarrow}r'_{b\uparrow}.
\end{eqnarray}
A similar formula can be written for the conductance $g(z)$:
\begin{equation}
g=\frac{e^2}{h}{\rm Tr \ } \frac{B(z)}{z |D(z)|^2},
\end{equation}
with:
\begin{eqnarray}
B(z)& = &|t'_{a\uparrow}|^2|t'_{b\uparrow}|^2\cos^2\frac{\theta}{2}
(1-zr_{a\downarrow}r'_{b\downarrow})(z-r_{a\downarrow}^*r'^*_{b\downarrow})
\nonumber \\
& +& |t'_{a\uparrow}|^2|t'_{b\downarrow}|^2\sin^2\frac{\theta}{2}
(1-zr_{a\downarrow}r'_{b\uparrow})(z-r_{a\downarrow}^*r'^*_{b\uparrow})
\nonumber \\
& + &|t'_{a\downarrow}|^2|t'_{b\uparrow}|^2\sin^2\frac{\theta}{2}
(1-zr_{a\uparrow}r'_{b\downarrow})(z-r_{a\uparrow}^*r'^*_{b\downarrow})
\nonumber \\
& + &|t'_{a\downarrow}|^2|t'_{b\downarrow}|^2\cos^2\frac{\theta}{2}
(1-zr_{a\uparrow}r'_{b\uparrow})(z-r_{a\uparrow}^*r'^*_{b\uparrow}).
\end{eqnarray}
Taking the average over the phases now amounts to coutour integration
for $z$:
\begin{equation}
\langle f \rangle=\frac{1}{2\pi i}
\oint \frac{dz}{z}\ \
f(z),
\end{equation}
where the integration is done along the unit circle. The result is
then given by the sum of the poles that are inside the unit circle. The
two poles of $D(z)$ are outside the unit circle, while the two poles
$z_1$ and $z_2$ of $z^2 D\left(\frac{1}{z}\right)$ are inside the
circle. They are given by:
\begin{eqnarray}
z_i& = &\frac{1}{2}\cos^2\frac{\theta}{2} (r_{a\uparrow}r'_{b\uparrow}
+ r_{a\downarrow} r'_{b\downarrow}) +\sin^2\frac{\theta}{2}
(r_{a\downarrow}r'_{b\uparrow}+r_{a\uparrow}r'_{b\downarrow})
\nonumber \\
& + &\frac{1}{2}(-1)^i
\left[
\cos^4\frac{\theta}{2} (r_{a\uparrow}r'_{b\uparrow}
- r_{a\downarrow} r'_{b\downarrow})^2 \right.
\nonumber \\
& + &
2\cos^2\frac{\theta}{2}\sin^2\frac{\theta}{2}
(r'_{b\downarrow} r'_{b\uparrow}(r_{a\uparrow} -r_{a\downarrow})^2
+r_{a\downarrow} r_{a\uparrow}(r'^2_{b\uparrow} + r_{a\downarrow}^2))
\nonumber \\
& + & \left.  \sin^4\frac{\theta}{2}
(r_{a\downarrow}r'_{b\uparrow}-r_{a\uparrow}r'_{b\downarrow})^2
\right]^{\frac{1}{2}}.
\end{eqnarray}
The averaged torque and conductance are then simply given by
\begin{equation}
\langle\frac{\partial\tau_b^x}{\partial V_0}\rangle=
- \frac{e\nu}{4\pi} \frac{\sin\theta}{z_1-z_2}
{\rm Tr \ }
\left( \frac{A(z_1)}{D(z_1)}-\frac{A(z_2)}{D(z_2)}\right)
\end{equation}
and
\begin{equation}
g=\frac{e^2}{h}\frac{1}{z_1-z_2}{\rm Tr \ } \left(\frac{B(z_1)}{D(z_1)}
-\frac{B(z_2)}{D(z_2)}\right).
\end{equation}
In the case where all the channels are not identical, these results
can be generalized  by introducing a $k$
dependence of the different transmission/reflection amplitudes.


\begin{thebibliography}{99}
\bibitem{deceptively} We use the term ``spin transfer'' to describe the
transfer, during the process of electron scattering, of spin angular
momentum between an electron current and a magnet.  The term should be used
with care in describing the current-induced interaction between adjacent
magnetic layers, because there is not a simple transfer of angular momentum
from one layer to the other -- spin may also flow away to infinity.  The
total magnetic moment of the two magnetic layers is therefore not
conserved;
 see our
discussion in section~\ref{1}.
\bibitem{noteslon}
In the notation of Ref.~\onlinecite{slon1}, the
fraction of the channels that act as perfect filters is
$J_b/(J_a+J_b)$.
\end{thebibliography}
\end{document}